\documentclass[a4paper,fleqn,usenatbib]{mnras}

\usepackage[T1]{fontenc}
\usepackage{ae,aecompl}

\usepackage{graphicx}	% Including figure files
\usepackage{amsmath}	% Advanced maths commands
\usepackage{amssymb}	% Extra maths symbols

\def\msun{\hbox{M$_\odot$}}
\title[A detached SMC stellar structure]{Revisiting a detached stellar structure in the
outer northeastern region of the Small Magellanic Cloud}

\author[Andr\'es E. Piatti]{
Andr\'es E. Piatti$^{1,2}$\thanks{E-mail: andres.piatti@unc.edu.ar} \\
$^{1}$Instituto Interdisciplinario de Ciencias B\'asicas (ICB), CONICET-UNCUYO, Padre J. Contreras 1300, M5502JMA, Mendoza, Argentina\\
$^{2}$Consejo Nacional de Investigaciones Cient\'{\i}ficas y T\'ecnicas, Godoy Cruz 2290, C1425FQB,  Buenos Aires, Argentina\\
}

\date{Accepted XXX. Received YYY; in original form ZZZ}

\pubyear{2021}

\begin{document}
\label{firstpage}
\pagerange{\pageref{firstpage}--\pageref{lastpage}}
\maketitle

\begin{abstract}
The outer northeastern region of the Small Magellanic Cloud (SMC) is populated by 
a shell-like overdensity whose
nature was recently investigated. We analyzed twenty catalogued 
star clusters projected onto it  from Survey of the MAgellanic Stellar History data sets. 
After carrying out a cleaning of field stars in the star cluster colour-magnitude diagrams
(CMDs), and deriving their astrophysical properties from the comparison between the
observed and synthetic CMDs, we found that four objects are not genuine star clusters, 
while the remaining ones are young star clusters (11, age $\sim$ 30-200 Myr) and
 intermediate-age (5, age $\sim$ 1.7-2.8 Gyr) star clusters, respectively. The resulting distances
show that intermediate-age and some young star clusters belong to the SMC 
main body, while the remaining young star clusters are nearly 13.0 kpc far away from
those in the SMC, revealing that the shell-like overdensity is more extended along
the line-of-sight than previously thought. We also found a clear age trend
and a blurred metallicity correlation along the line-of-sight of young clusters, in
the sense that the farther a star cluster from the SMC, the younger, the more
metal rich, and the less massive it is. These young clusters are also affected
by a slightly larger interstellar reddening than the older ones in the shell-like
overdensity. These outcomes suggest that the shell-like overdensity can
possibly be another tidally perturbed/formed SMC stellar structure
from gas striped off its body, caused by the interaction with the Large Magellanic
Cloud or the Milky Way.
\end{abstract} 


%%%%%%%%%%%%%%%%%%%%%%%%%%%%%%%%%%%%%%%%%%%%%%%%%%

%%%%%%%%%%%%%%%%% BODY OF PAPER %%%%%%%%%%%%%%%%%

\section{Introduction}

\citet{martinezdelgadoetal2019} studied the nature of a stellar structure located toward the
outer northeastern region of the Small Magellanic Cloud (SMC), which they referred to as a 
shell-like overdensity. By using the Survey of the MAgellanic Stellar History 
\citep[SMASH,][]{nideveretal2017} and the second release (DR2) of the {\it Gaia} mission \citep{gaiaetal2016,gaiaetal2018b} data sets, they inferred that the shell-like overdensity
is composed by stars younger than $\sim$ 150 Myr. As far as its origin is considered, 
they concluded that it formed from a recent star formation even, possible caused by
the interaction of the SMC with the Large Magellanic Cloud (LMC) or the Milky Way (MW), and
discarded the possibilities of being a tidally disrupted stellar system or a bright part of a 
spiral arm-like structure.  From available ages of nine star clusters projected onto the shell-like 
overdensity, they confirmed the youth of such a stellar structure. 

The  resulting stellar content and structure of the shell-like overdensity 
are based on the analysis of colour-magnitude diagrams (CMDs), assuming a
mean distance modulus of 18.96 mag. The same fixed distance modulus was also
assumed for the nine studied star clusters. However, the SMC is known to be more
extended along the line-of-sight than  the size of the galaxy projected in the sky
\citep{ripepietal2017,muravevaetal2018,graczyketal2020}, so
that by adopting a mean distance for every single star and star cluster could
mislead the interpretation about the structure, stellar population and the
possible origin of this SMC stellar structure. As far as we are aware, accurate distances 
of star clusters have been previously measured for a limited number of objects 
\citep[19 star clusters; ][]{cetal01,glattetal2008a,diasetal2016,mvetal2021,diasetal2021}. 
Recently, \citet{piatti2021c} considered the SMC as a triaxial
spheroid, and estimated deprojected distances for the SMC star clusters catalogued by \citet{bicaetal2020}. 
By adopting a 3D geometry of the SMC, he avoided the spurious effects 
caused by assuming that a star cluster observed along the line-of-sight is close to the 
galactic centre.

Precisely, the main aim of this work is to provide for the first time accurate distances, ages 
and metallicities for all the star clusters projected onto the shell-like overdensity, thus
mitigating the lack of homogeneous estimates of these astrophysical properties. 
Fortunately, this scope is nowadays possible to be achieved because of the availability
of fairly deep photometry and sophisticated statistical methods to fit theoretical isochrones
to the star clusters' CMDs. An additional and not negligible aspect of this analysis comprises
the appropriate decontamination of field stars from the star clusters' CMDs. The combined
star cluster CMD built by \citet[][see their figure 8]{martinezdelgadoetal2019} shows mainly
the contribution of main sequence stars of massive young star clusters down to $i$ $\sim$
22 mag and of fainter stars which belong to the composite SMC star field.
Both cluster and field stars may populate the observed combined subgiant, red giant 
branches and red clumps, so that intermediate-age and young clusters should not be ruled out. 
For this reason, a field star cleaning of the observed star clusters'
CMDs is necessary before estimating their ages, distances and metallicities. 

Once cleaned star clusters'  CMDs are obtained, it is possible to investigate their 3D
spatial distribution, and hence to draw conclusions about the location and dimensions
of the shell-like overdensity, provided that they belong to that stellar structure. Their
spatial distribution can also hint at its possible origin, namely, as a stellar structure that 
formed recently in the outskirts of the SMC, a detachment of some part of the SMC, or
both mechanisms combined. Bearing in mind the above considerations, we describe in 
Section 2 the gathered star cluster sample, the publicly available retrieved data sets, and the
cleaning procedure applied in order to build star clusters' CMDs with highly probable
members. Section 3 deals with the cleaned CMDs and cluster properties estimates, while
in Section 4 we analyze and discuss the nature of the shell-like overdensity. The
main conclusions of this work are given in Section 5.

\begin{figure}
\includegraphics[width=\columnwidth]{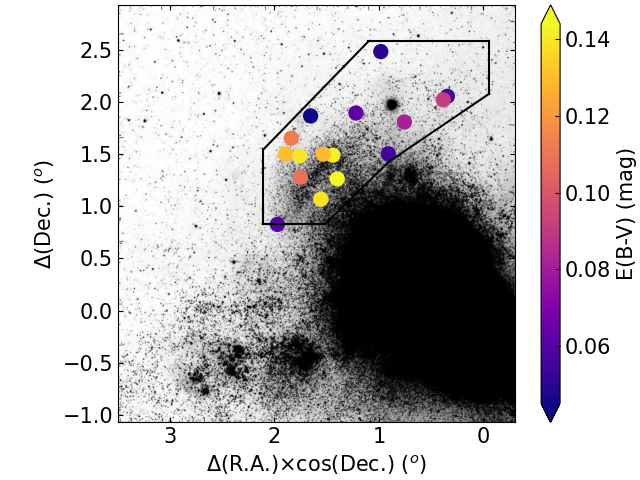}
\caption{Coloured symbols represent star clusters located within the shell-like overdensity
as delimited by \citet[][Fig.~8]{martinezdelgadoetal2019}, overplotted over their
Fig. 5.}
\label{fig1}
\end{figure}

\section{Data handling}

We started by searching the star cluster catalogue compiled by \citet{bicaetal2020} any
star cluster located within the region of the shell-like overdensity delimited by 
\citet{martinezdelgadoetal2019}. Particularly, we used as reference the boundaries of their 
figure 8 (upper
panel), which are also illustrated in Fig.~\ref{fig1}. We found 20 catalogued
star clusters, among them the nine star clusters used by \citet[][see their Table 1]{martinezdelgadoetal2019}.
Four catalogued star clusters (OGLE-SMC~274, 275, 276, and 276) were not confirmed
as genuine physical systems in our field star decontamination procedure, so that we
discarded them from our analysis. The remaining catalogued star clusters are listed
in Table~\ref{tab1}. 

We used the Survey of the Magellanic Stellar History
(SMASH) DR2 data sets \citep{nideveretal2021} to conduct the analysis of the selected star clusters.
Coordinates (R.A. and Dec.), PSF $g,i$ magnitudes and their respective errors, $\chi$ and 
{\sc sharpness} parameters of stellar sources
located inside a radius of 9$\arcmin$ from the star clusters' centres listed by \citet{bicaetal2020} were
retrieved from the portal of the Astro Data Lab\footnote{https://datalab.noao.edu/smash/smash.php}, 
which is part of the Community Science and Data 
Center of NSF’s National Optical Infrared Astronomy Research Laboratory.
The retrieved data sets consist of sources with 0.2 $\le$ {\sc sharpness} $\le$ 1.0  and
$\chi^2$ $<$ 0.5, so bad pixels, cosmic rays, galaxies, and unrecognized double 
stars were excluded. According to \citet{bicaetal2020}, the radii of the selected star clusters are 
relatively small, from 0.35$\arcmin$ up to  1.7$\arcmin$, with an average of 0.9$\arcmin$  
(NGC~458 is the largest cluster in the sample with a radius of 2.6$\arcmin$).
Because we downloaded information
for circular areas much larger than the star clusters' radii, we thoroughly monitored 
the  contamination of field stars in the star clusters' CMDs. 

When dealing with star cluster CMDs, it is necessary to take into account the contamination
of field stars, because it is not straightforward  to consider a star as a cluster member only
on the basis of  its position in that CMD. Frequently it is assumed that the colours and
magnitudes of field stars projected onto the star cluster field are similar to those around the
star cluster. However, even though a star cluster is not projected onto a 
crowded star field or is not affected by differential reddening, it is highly possible to 
find different magnitude and colour distributions of field stars located along the star cluster
line-of-sight, and in the surrounding field as well. For this reason, it is convenient to use field stars
located all around the star cluster. 

The field star decontamination procedure adopted in this work is based on that devised by 
\citet{pb12}, which was satisfactorily applied in cleaning CMDs of star clusters 
projected toward crowded star fields  \citep[e.g.,][and references therein]{p17a} 
and affected by differential  reddening \citep[e.g.,][and references therein]{p2018}. 
We refer the reader to the recent works by \citet{piatti2021b} and \citet{piatti2021d}, 
where details of the cleaning technique are provided. The method
comprises three main steps: 1) to properly reproduce the distribution of magnitudes and
colours of stars in the surroundings of the star cluster; 2) to reliably subtract
those magnitude and colour distributions from the star cluster CMD and; 3) to assign membership 
probabilities to stars that survived the cleaning procedure.
Stars with relatively high membership probabilities can likely be cluster members,
provided that they are placed along the expected CMD star cluster sequences and belong to a
stellar overdensity.

The sky regions used to map the field star density, and the distribution of magnitudes 
and colours of field stars consist of six circles of radius 3$\arcmin$ distributed 
uniformly around the  star cluster circle, which is also of the same radius. These surrounding 
circles are thought to be placed far from the star cluster, but not too far from it as to 
become unsuitable as representative of the star field projected along the
line-of-sight of the star cluster. We cleaned a star cluster CMD six times.
Each execution used the observed star cluster CMD and one of the six associated surrounding 
fields, and produced a cleaned star cluster CMD. Thus, we obtained six different cleaned 
CMDs for each star cluster. During each individual run, we subtracted from a star cluster CMD 
a number  of stars equal to that in the chosen surrounding field. In order to subtract
stars with the same magnitude and colour distributions of the field stars, we
defined boxes centred on the magnitude and colour of each star of the surrounding field CMD,
then superimposed them on the star cluster CMD, and finally chose one star per box to subtract.
In the present work, we used initial boxes of ($\Delta$$g$, $\Delta$$(g-i)$) =
(1.0 mag,0.25 mag) centred on the ($g$, $g-i$) values of each field star. 
In the case that more than one star is located inside that delimited CMD region, the 
closest one to the centre of that (magnitude, colour) box is subtracted. 

We finally used the six different cleaned star cluster CMDs to assign membership
probabilities to surviving stars, as follows:  $P$ ($\%$) = 100$\times$$N$/6, where $N$ 
represents the number of times a star was not subtracted during the six different CMD cleaning 
executions. Hence, a star with $P$ = 100 is a star not subtracted during any of the cleaning runs
that involved six different surrounding fields, separately. For this reason, it has the highest chance 
to contribute to the intrinsic features of the cleaned CMD. A star with P = 16.67 is that that survived
once from six different cleaning executions, meaning that its magnitude and colour were mostly found 
in the surrounding field population. With that
information on hand, we built Fig.~\ref{fig2}, which shows the spatial distribution
and the CMD of all the measured stars of the selected star cluster sample.

%\begin{flushleft}
\begin{table*}
\caption{Derived astrophysical properties of selected SMC star clusters}
\begin{tabular}{@{}lccccccccc}\hline
Star cluster & E(B-V) & $m-M_o$ & log(age /yr) & [Fe/H] & Mass & binarity \\
                   & (mag)	& (mag)	&	& (dex) &	($\msun$) & \\\hline

B88	&	0.056$\pm$0.030&	18.998$\pm$0.096&	8.131$\pm$0.159&	0.0086$\pm$0.0029&	622$\pm$190&		0.27$\pm$0.14\\
B111	&	0.081$\pm$0.048&	18.868$\pm$0.153&	8.271$\pm$0.148&	0.0119$\pm$0.0022&	212$\pm$91&		0.30$\pm$0.14\\
B139	&	0.144$\pm$0.035&	19.098$\pm$0.111&	7.812$\pm$0.035&	0.0076$\pm$0.0042&	1494$\pm$398&	0.32$\pm$0.12	\\
BS~116 &	0.055$\pm$0.039&	18.749$\pm$0.124&	9.274$\pm$0.070&	0.0026$\pm$0.0007&	1740$\pm$417&	0.29$\pm$0.13 \\
H86-197& 0.112$\pm$0.034&	18.798$\pm$0.108&	7.966$\pm$0.440&	0.0106$\pm$0.0024&	371$\pm$171&		0.28$\pm$0.14 \\
HW~33 &	0.091$\pm$0.035&	18.956$\pm$0.111&	7.918$\pm$0.365&	0.0083$\pm$0.0041&	1057$\pm$311&	0.26$\pm$0.14	\\
HW~56&	0.062$\pm$0.039&	18.749$\pm$0.124&	9.442$\pm$0.060&	0.0022$\pm$0.0008&	2251$\pm$722&	0.30$\pm$0.13 \\
HW~64&	0.142$\pm$0.026&	19.193$\pm$0.083&	7.468$\pm$0.309&	0.0064$\pm$0.0029&	1046$\pm$288&	0.29$\pm$0.14 \\
HW~67&	0.045$\pm$0.040&	18.709$\pm$0.127&	9.276$\pm$0.046&	0.0027$\pm$0.0013&	2189$\pm$563&	0.36$\pm$0.11	\\
HW~73&	 0.130$\pm$0.014&	19.180$\pm$0.045&	7.868$\pm$0.085&	0.0073$\pm$0.0037&	1621$\pm$512&	0.25$\pm$0.14 \\
IC~1655&	0.129$\pm$0.012&	18.659$\pm$0.039&	8.054$\pm$0.107&	0.0037$\pm$0.0015&	4752$\pm$180&	0.17$\pm$0.06	\\
IC~1660&	0.139$\pm$0.017&	18.813$\pm$0.055&	7.909$\pm$0.143&	0.0054$\pm$0.0023&	3960$\pm$497&	0.15$\pm$0.09 \\
L73	&	0.050$\pm$0.041&	18.829$\pm$0.131&	9.269$\pm$0.036&	0.0034$\pm$0.0010&	3716$\pm$676&	0.27$\pm$0.11	\\
L95	&	0.140$\pm$0.030&	19.176$\pm$0.096&	7.585$\pm$0.372&	0.0067$\pm$0.0041&	910$\pm$326&		0.24$\pm$0.15	\\
L100	&	0.060$\pm$0.031&	18.781$\pm$0.099&	9.417$\pm$0.049&	0.0035$\pm$0.0005&	4649$\pm$272&	0.11$\pm$0.07 \\
NGC~458&0.109$\pm$0.010&	18.736$\pm$0.033&	8.130$\pm$0.070&	0.0062$\pm$0.0011&	4853$\pm$126&	0.06$\pm$0.04	\\
\hline
\end{tabular}
\label{tab1}
\end{table*}
%\end{flushleft}

\begin{figure}
\includegraphics[width=\columnwidth]{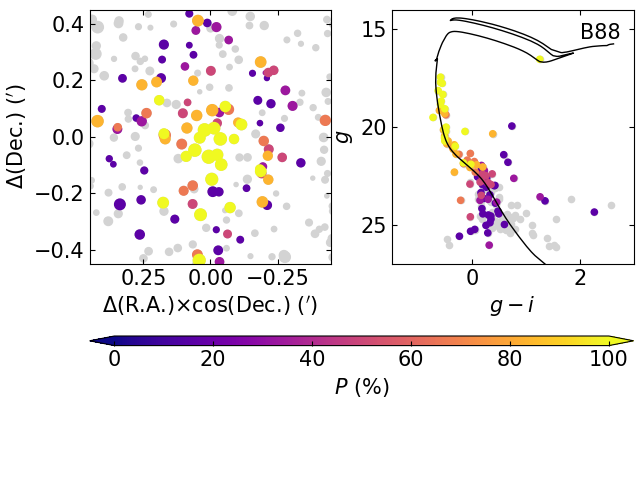}
\includegraphics[width=\columnwidth]{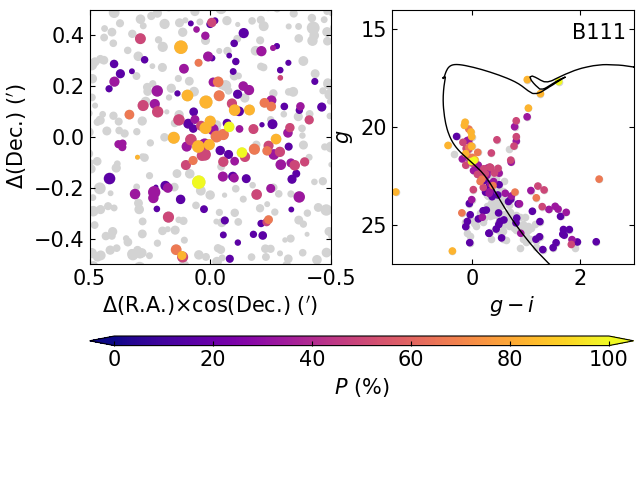}
\includegraphics[width=\columnwidth]{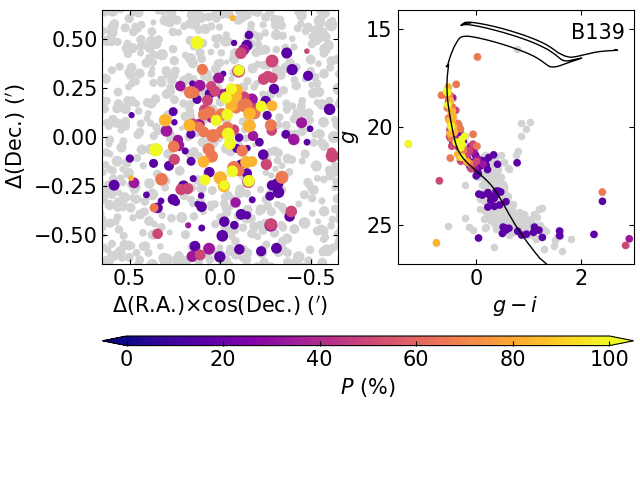}
\caption{Charts of the star clusters (left panel), and
the respective CMDs (right panel), drawn with grey symbols. 
The size of the charts' symbols is proportional to the $g$ brightness of the star.
Coloured symbols in both panels represent the
stars that survived the CMD cleaning procedure, colour-coded
according to the assigned membership probabilities ($P$).  A theoretical
isochrone \citep{betal12} for the mean derived parameters is
superimposed onto the CMD with a black line. }
\label{fig2}
\end{figure}

\setcounter{figure}{1}
\begin{figure}
\includegraphics[width=\columnwidth]{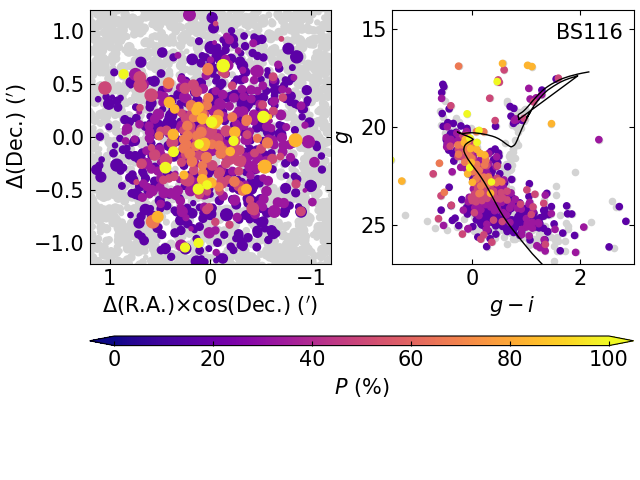}
\includegraphics[width=\columnwidth]{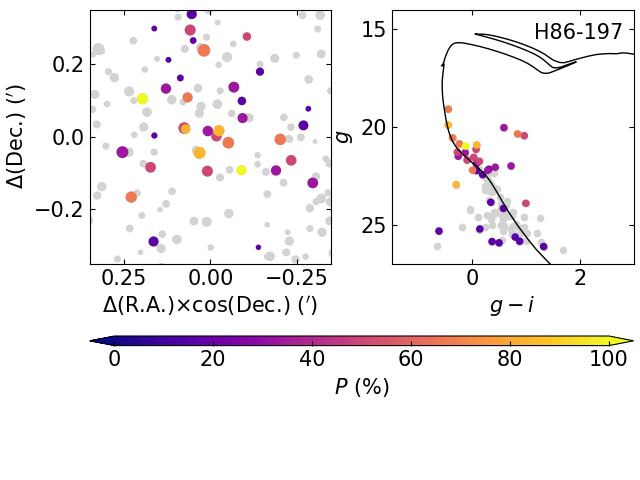}
\includegraphics[width=\columnwidth]{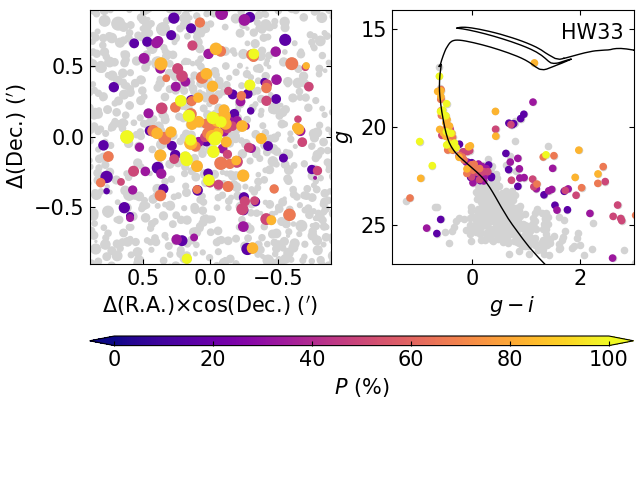}
\caption{continued.}
\end{figure}

\setcounter{figure}{1}
\begin{figure}
\includegraphics[width=\columnwidth]{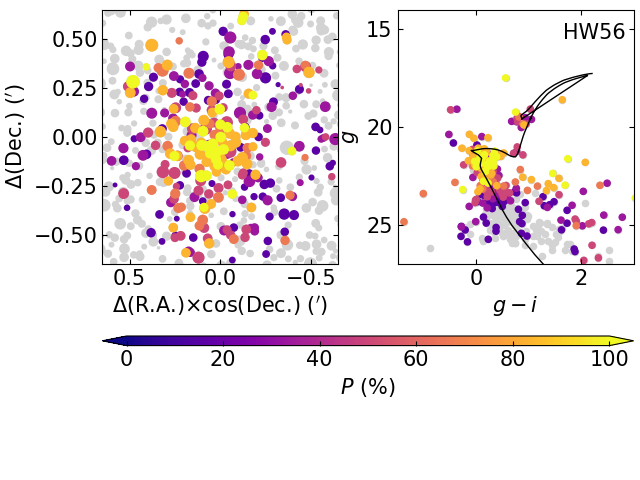}
\includegraphics[width=\columnwidth]{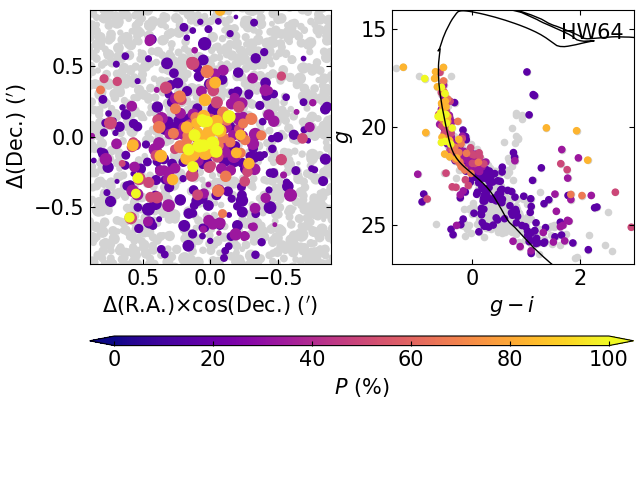}
\includegraphics[width=\columnwidth]{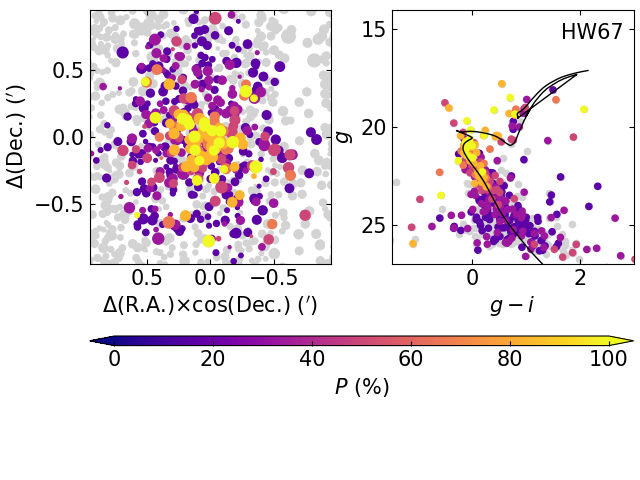}
\caption{continued.}
\end{figure}

\setcounter{figure}{1}
\begin{figure}
\includegraphics[width=\columnwidth]{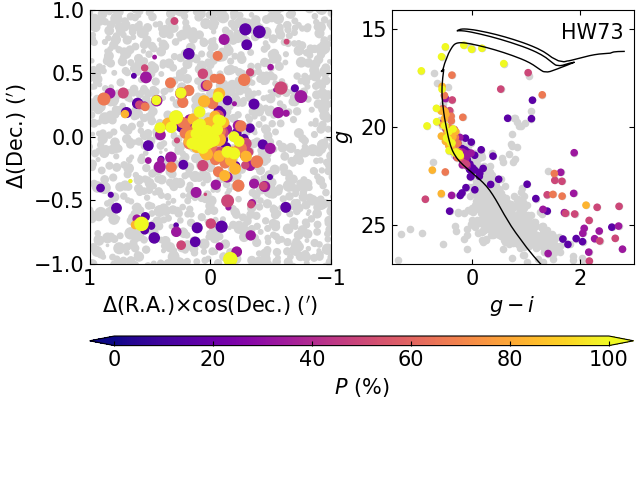}
\includegraphics[width=\columnwidth]{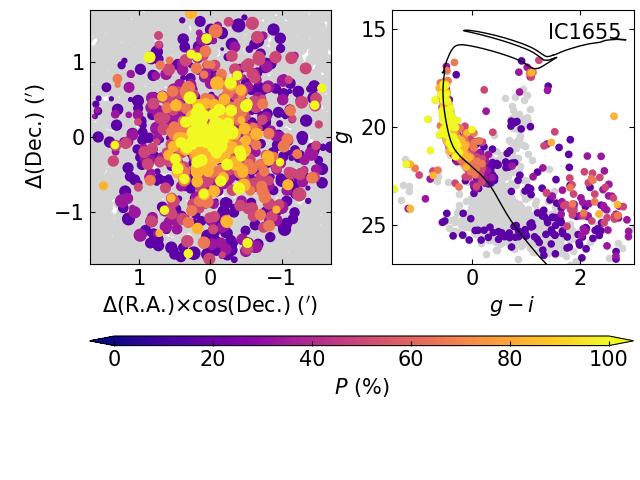}
\includegraphics[width=\columnwidth]{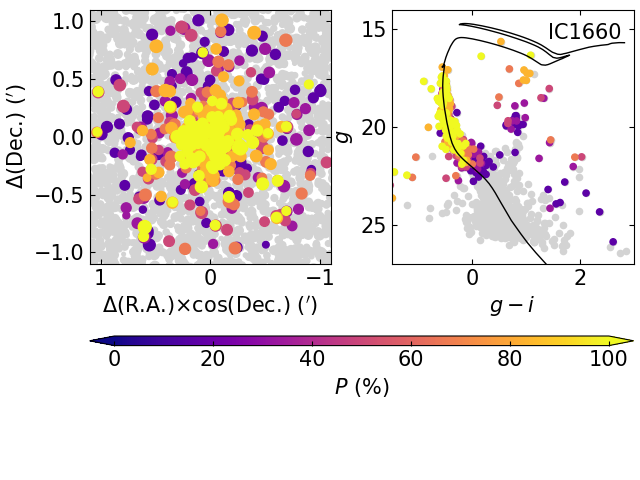}
\caption{continued.}
\end{figure}

\setcounter{figure}{1}
\begin{figure}
\includegraphics[width=\columnwidth]{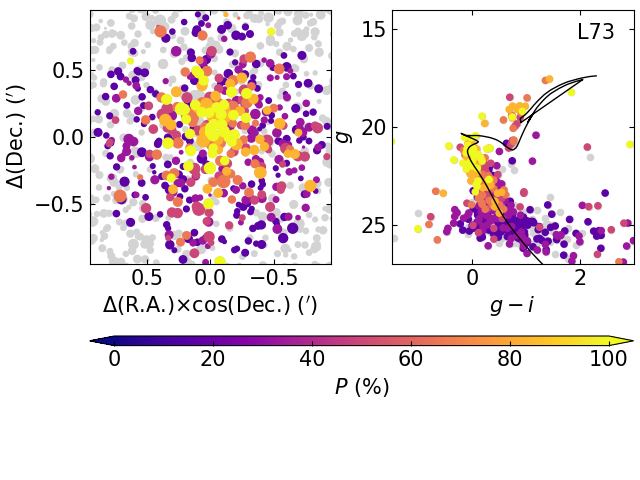}
\includegraphics[width=\columnwidth]{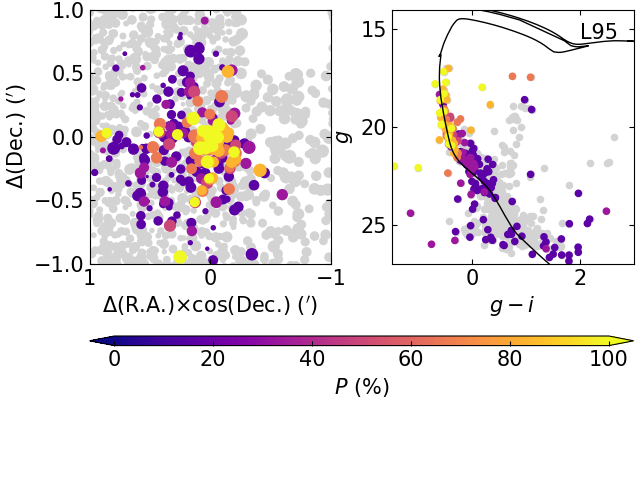}
\includegraphics[width=\columnwidth]{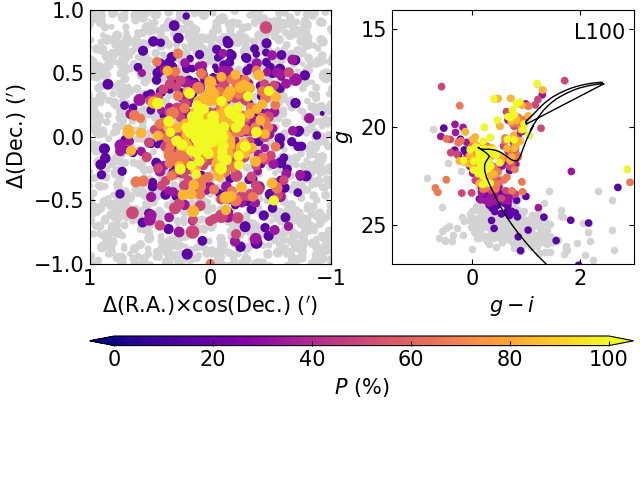}
\caption{continued.}
\end{figure}

\setcounter{figure}{1}
\begin{figure}
\includegraphics[width=\columnwidth]{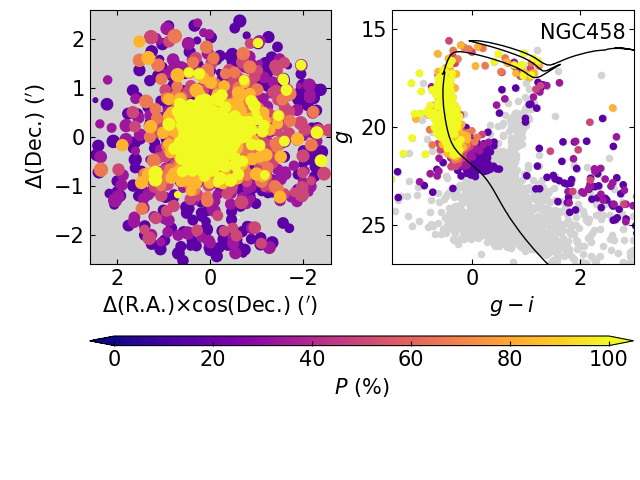}
\caption{continued.}
\end{figure}

\section{Star cluster parameters}

The resulting cleaned CMDs highlight the star cluster sequences on top
of those of field stars. As can be seen, there seem to be moderately old
star clusters projected onto younger star fields (e.g., BS~116, L110);
young clusters projected onto older fields (e.g., L95, NGC~458); young
star clusters projected onto nearly young star fields (e.g., B88, H86-197);
and relatively old star clusters projected onto star fields of similar age
(e.g., HW~56, L73). A common feature seen in  the cleaned CMDs
is the presence of field stars ($P<$ 50$\%$) at the fainter part of the Main 
Sequence, as expected. Because of stochastic effects some 
residuals at any $P$ value also remain, which confirm that the
assumption of a uniform surrounding star field is not appropriate.

We used the cleaned star cluster CMDs to derive accurate star cluster parameters
by employing routines of the Automated Stellar Cluster Analysis code \citep[\texttt{ASteCA,}][]{pvp15}
that allowed us to derive all of them simultaneously. \texttt{ASteCA} is a suit of tools designed to analyze 
data sets of star clusters in order to determine their basic properties. We thus 
obtained a synthetic CMD that best matches the cleaned star cluster CMD. The metallicity, age, 
distance,  reddening, star  cluster present mass and binary fraction associated to that 
generated synthetic CMD were adopted as the best-fitted star cluster properties. 

We started by using the theoretical isochrones computed by \citet{betal12} for the SMASH photometric
system. We downloaded theoretical isochrones for different metallicities values, from $Z$ = 0.000152
([Fe/H]=-2.0 dex) up to 0.030152 ([Fe/H]=0.30 dex) in steps of $\Delta Z$=0.001. This metallicity
range covers almost all the metallicity regime of the Magellanic Clouds \citep{pg13}. This is an
important consideration, because the studied star clusters lie in an outer region of the SMC,
where  metal-poor old and metal-rich young objects formed at the galaxy formation and 
galaxy interaction, respectively. As for ages,
we downloaded isochrones from log(age /yr)=6.0 (1 Myr) up to 10.1 (12.5 Gyr) in steps of
$\Delta$log(age /yr)=0.025. In total, we gathered nearly 7800 different theoretical isochrones.

The steps by which a synthetic star cluster for a given set of age, metallicity, distance modulus, and
 reddening values is generated by \texttt{ASteCA} is as follows: i) a theoretical isochrone
is picked up, densely interpolated to contain a thousand points throughout its entire length,
including the most evolved stellar phases. ii) The isochrone is shifted in colour and
magnitude according to the $E(B-V)$ and $(m-M)_o$ values to emulate the effects these
parameters have over the isochrone in the CMD. 
iii) The isochrone is trimmed down to a certain faintest magnitude 
according to the limiting magnitude derived from the CMD. iv) An initial mass function 
(IMF) is sampled in the mass range $[{\sim}0.01{-}100]\,M_{\odot}$ up
to a total mass value $M$ provided that 
evolved CMD regions result properly populated.
The distribution of masses is then used to obtain a properly populated synthetic 
CMD by keeping one star in the interpolated 
isochrone for each mass value in the distribution. v) A random fraction of stars is 
assumed to be binaries, which is set by default to  
$50\%$ \citep{von_Hippel_2005}, with secondary masses 
drawn from a uniform distribution between the mass of the primary star and a 
fraction of it given by a mass ratio parameter set to $0.5$. 
vi) An appropriate
magnitude completeness and an exponential photometric error functions are
finally applied to the synthetic star cluster. 

The input data sets consist of the cleaned $g,i$ photometry with their respective uncertainties,
alongside the associated membership probabilities, i.e., all coloured points in Fig.~\ref{fig2}.
For generating the synthetic CMDs, we adopted the initial mass function of \citet{kroupa02}; 
a minimum mass ratio for the generation of binaries of 0.5; and a range of true distance moduli 
from 18.5 mag (50 kpc) up to 19.5 mag (80 kpc). Star cluster mass and binary fractions
were set in the ranges 100-5000 $\msun$ and 0.0-0.5, respectively. 
We explored the parameter space of the 
synthetic CMDs through the minimization of the likelihood function defined by 
\citet[][; the Poisson likelihood ratio (eq. 10)]{tremmeletal2013} using a parallel tempering 
Bayesian MCMC algorithm, and the optimal binning \citet{knuth2018}'s method.
Errors in the obtained parameters are
estimated from the standard bootstrap method described in \citet{efron1982}. We refer the reader
to the work of \citet{pvp15} for details concerning the implementation of these algorithms.
Table~\ref{tab1} lists the resulting parameters for the studied star cluster sample. We 
illustrate the performance of the parameter matching procedure by superimposing the isochrone 
corresponding to the best-fitted synthetic CMD to the cleaned star cluster CMDs (see
Fig.~\ref{fig2}).

\section{Analysis and discussion}

By inspecting Table~\ref{tab1} we found that more than 30$\%$ of the star clusters turned out
to be older than 1.7 Gyr, while the remaining ones resulted to be younger than 150 Myr. We
confirmed that the nine star clusters used by \citet[][see their Table 1]{martinezdelgadoetal2019}
to study the nature of the shell-like overdensity
are young star clusters, to which we added two other young star clusters (B111, H86.197). Note that
the ages compiled by them come from a variety of sources, while our age estimates were
obtained from  SMASH data sets and the procedures described in 
Sections 2 and 3, namely: 
the decontamination of field stars and the comparison between observed and synthetic CMDs.
For these reasons, we think that our derived star cluster properties are homogeneous and 
more accurate. They come from deeper photometric data; they were obtaining without assuming 
a mean distance modulus for all the star clusters, and the associated synthetic CMDs best resemble 
the observed ones among a large number of generated CMDs covering the parameter space.

\begin{figure}
\includegraphics[width=\columnwidth]{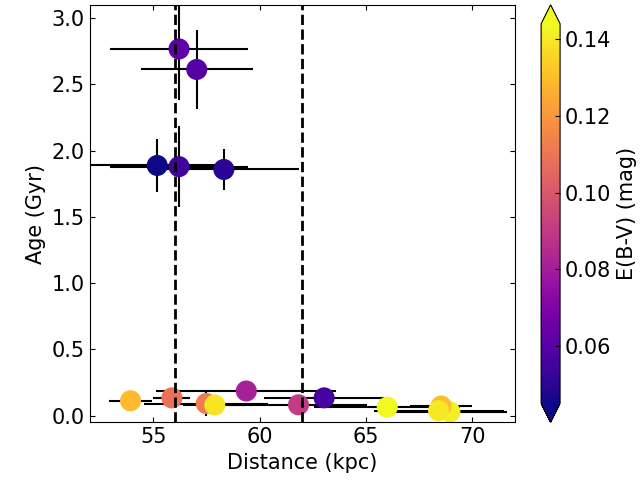}
\caption{Relation between different star cluster properties. The vertical dashed lines
represent the boundaries of the SMC main body along its line-of-sight \citep{piatti2021a}.}
\label{fig3}
\end{figure}

As Fig.~\ref{fig1} shows, the resulting interstellar reddening values show that the studied clusters are
split into two groups with slightly different mean colour excesses $E(B-V)$, that in turn 
have some particular projected spatial distribution in the sky.
It seems that there is a group of mostly outermost star clusters with a mean $E(B-V)$ value 
of 0.05 mag, and another group of star clusters closer to the SMC main body with a mean
$E(B-V)$ value of 0.12 mag. Both mean $E(B-V)$ colour excesses are within the range of
values for that part of the SMC according to different SMC reddening maps 
\citep[see, e.g.,][]{belletal2020,skowronetal2021}. With the aim of looking into such a
spatial distribution we built Fig.~\ref{fig3}, that shows studied older star clusters belonging
to the group of objects with relatively small $E(B-V)$ values and most of the younger
star clusters being affected by larger interstellar reddening values. Younger star clusters
are frequently found in regions with  gas and dust residuals, so that Figs.~\ref{fig1}
and \ref{fig3} confirm that a region of the shell-like overdensity contains young star 
clusters and gas/dust residuals \citep[see, also, ][]{martinezdelgadoetal2019}.

Fig.~\ref{fig3} also reveals a striking distribution of star clusters along the line-of-sight.
Older star clusters are located at distances compatible with being part of the SMC main body,
whose boundaries are nearly at 56 and 62 kpc from the Sun \citep[][and references therein]{piatti2021a}.
However, younger star clusters span a wider range of heliocentric distances, from those located in
front of the SMC ($D$ $\la$ 56 kpc) up to those placed behind it  ($D$ $\ga$ 62 kpc).
Such a range in heliocentric distance is also seen in SMC stellar structures
that formed from the interaction with the LMC. For instance, the Magellanic Bridge is located
at $D$ $<$ 55 kpc \citep{wagnerkaiser17,jacyszyndobrzenieckaetal2020}; the Counter-Bridge,
placed towards the northeastern outskirts of the SMC, is at $D$ $>$ 65 kpc \citep{diasetal2021},
and the West halo, in the south-west SMC periphery, is at 52 $\la$ $D$ (kpc) $\la$ 70 
\citep{diasetal2016}.

\begin{figure}
\includegraphics[width=\columnwidth]{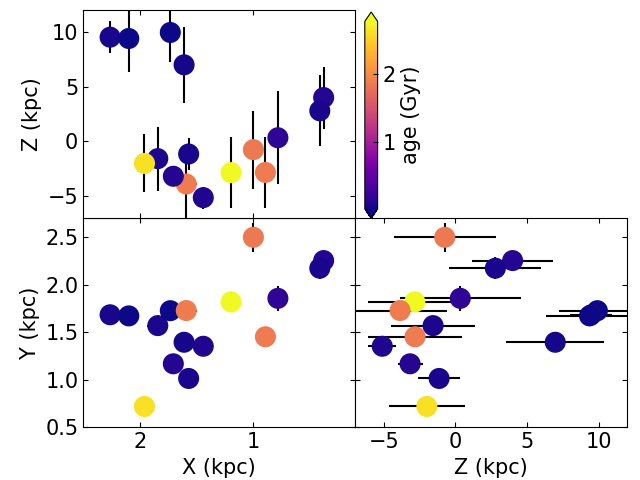}
\caption{3D spatial distribution of SMC star clusters in the shell-like overdensity. Errobars
are included; for $X$ and $Y$ coordinates they are smaller than the size of the symbols.}
\label{fig4}
\end{figure}

We transformed star cluster equatorial coordinates and heliocentric distances to 3D cartesian 
coordinates following the formalism described in \citep{wn2001,vdmarel2001a}, and the SMC
optical centre coordinates R.A. = 13.1875$\degr$, Dec. = -72.8286$\degr$ \citep{cetal01}, and
its heliocentric distance of  59.0 kpc \citep{piatti2021a}. Fig.~\ref{fig4} depicts the resulting
$X,Y,Z$ coordinates relative to the SMC centre. The $Z$ direction is perpendicular to the
($X,Y$) plane and is oriented along the line-of-sight to the SMC centre; it increases outwards.
The ($X$,$Y$) plane, which is a fairly close representation of the distribution of the star
clusters seen in the sky (see Fig.~\ref{fig1}), shows that the older star clusters are 
scattered across the shell-like overdensity. When looking at their distribution along the Z axis, 
we found that all of them belong to the SMC main body. There are also other seven younger
star clusters pertaining to the galaxy that are mingled with the older star clusters. From this
finding, we conclude that the shell-like overdensity could not purely be originated by a recent 
star formation event. These younger star clusters could formed in that part of the galaxy
when the studied older star clusters also populated that outermost SMC region.

\begin{figure}
\includegraphics[width=\columnwidth]{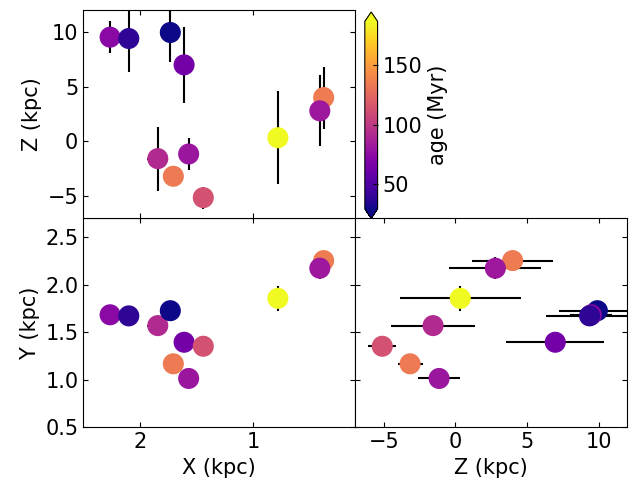}
\caption{Same as Fig.~\ref{fig4} for younger star clusters.}
\label{fig5}
\end{figure}

The possible formation scenario of the younger star clusters can be unveiled by including 
those star clusters located far away the SMC main body. There are eight young star 
clusters projected towards the southeast region of the shell-like overdensity, four of them placed 
nearly 13.0 kpc far away from those in the SMC. Fig.~\ref{fig5} shows the
3D spatial distribution of the studied younger star clusters. It reveals that, for the southeastern
group of young star clusters, the younger the
star cluster, the larger its $Z$ coordinate. This range of $Z$ values resembles that 
of tidally perturber/formed  SMC stellar structures mentioned above. Additionally, the age gradient
seen along the $Z$ direction among the group of southeastern young star clusters tells us about
a possible cluster formation wave that started nearly 200 Myr ago (the age of B111) at the 
outermost region of the SMC and continued outwards until very recently, $\sim$ 30 Myr ago 
(the age of HW~64). In order to that
cluster formation sequence took place in space and time, gas from the SMC should have
stripped off its body, so that while gas is moving outwards, star clusters formed.
Here we think that the interaction with the LMC/MW \citep{beslaetal07,beslaetal2012} could
be responsible for the shell-like overdensity is another possible tidally perturbed/formed SMC
stellar structure. This outcomes bring additional support to previous speculations and results
of this possible formation mechanism \citep[][and references therein]{getal10,setal14,jacyszynetal2017,ripepietal2017,zivicketal2018,zivicketal2019}.

\begin{figure}
\includegraphics[width=\columnwidth]{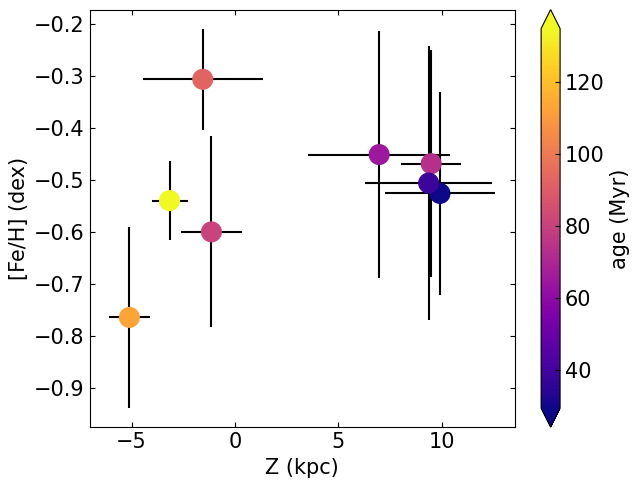}
\caption{Relation between the metallicity and the relative depth from the SMC centre
of the group of younger star clusters  located in the southeastern region of
the shell-like overdensity.}
\label{fig6}
\end{figure}

\begin{figure}
\includegraphics[width=\columnwidth]{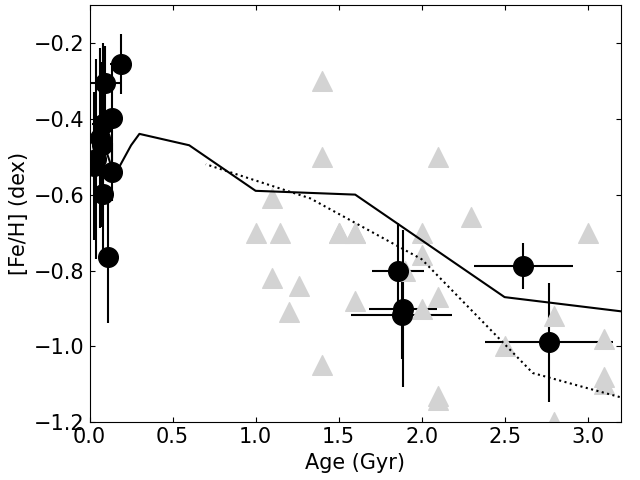}
\caption{The age-metallicity relationship of shell-like overdensity star clusters.
Triangles represent the SMC star clusters from \citet{pg13}, while solid and dotted
lines correspond to the age-metallicity relationships derived from field stars
\citep{hz04}, and theoretical models \citep{pt1008}, respectively. }
\label{fig7}
\end{figure}

Fig.~\ref{fig6} shows a slight correlation between the metallicity and the relative depth
along the line-of-sight to the SMC centre of the eight young star clusters in the southeastern 
region of the
shell-like overdensity. Such a trend could be interpreted as a chemical abundance enrichment
process with time along the $Z$ direction. Nevertheless, further investigations from 
more accurate metallicities are needed. It is also worth mentioning that among these
eight young star clusters the more massive (total present mass $\sim$ 4500 $\msun$) ones
formed in the SMC main body, while those less massive ($<$ 2000 $\msun$) are located
farther away from the SMC. As far as the age-metallicity relationship of
SMC star clusters is considered, Fig.~\ref{fig7} shows that the studied star clusters
follow the known trend and dispersion derived from observed star clusters
\citep[see][]{pg13}, field stars \citep{hz04}, and theoretical models \citep{pt1998},
respectively.

\section{Conclusions}

We studied twenty catalogued star clusters projected onto the so-called shell-like
overdensity, a stellar structure located in the outer northeastern  region of the SMC, 
with the aim of estimating accurate star cluster properties. We embarked on this
investigation motivated by the fact the star clusters analysed by \citet{martinezdelgadoetal2019}
were assumed to be at the same mean SMC distance, while the galaxy is known
to be more extended in depth than its projection in the plane of the sky. An additional reason
supporting this study is the need of decontamination of field stars from the star
cluster CMDs in order to derive accurate astrophysical parameters. Ages and metallicities
were previously derived for some of these star clusters from different sources, so that an
effort of determining them homogeneously is also valuable.

A robust field star cleaning procedure was applied to the entire star cluster sample using
SMASH data sets, which unveiled that four objects are not
genuine star clusters (OGLE-SMC~274, 275, 276, and 277), while the remaining true 
star clusters are split in two age groups: a younger group (30-200 Myr) with eleven star clusters,
and an older one (1.7-2.8 Gyr) composed of five star clusters. We confirmed the youth 
of the nive star clusters used by \citet{martinezdelgadoetal2019}. When star cluster
distances are examined, we found that the older star clusters and seven young clusters
pertain to the SMC main body, while four young clusters are nearly 13.0 kpc farther away.
This outcome suggests a more extended shell-like overdensity structure along the line-of-sight 
than in the plane of the sky. The younger star clusters that span a wide range of depth values
are located in the southeastern region of the shell-like overdensity. Curiously, they are
affected by slightly larger colour excesses as compared with the other star clusters in
the shell-like overdensity. There is also a correlation between their ages and their distances
along the lne-of.sight to the SMC centre, in the sense that younger clusters ($\sim$ 30 Myr)
are farther than those relatively older ($\sim$ 200 Myr). A less clear correlation with 
metallicity also arises, with more metal-rich young star clusters placed farther from the
SMC.

These observational evidence point to the possibility that the younger star clusters
in the shell-like overdensity are part of a tidally perturbed/formed SMC stellar structure
from gas striped off its body, caused by the interaction with the LMC/MW. At the moment
of that interaction intermediate-age star clusters populated the shell-like overdensity region, 
while star clusters formed out of gas clouds pulled out by the interacting galaxy. The
age-metallicity relationship of the studied star clusters are in agreement with our
previous knowledge of age-metallicity relationships built from observed star clusters, 
star fields, and theoretical models.

The present results can stimulate further deep observational campaigns of SMC
star clusters. Indeed, we used {\it Gaia} 
EDR3\footnote{https://archives.esac.esa.int/gaia.} data sets to performe the analysis 
described in Section 2, by extracting
parallaxes ($\varpi$), proper motions in right ascension (pmra) and declination (pmdec),
excess noise (\texttt{epsi}), the significance of excess of noise (\texttt{sepsi}), and $G$,
$BP$, and $RP$ magnitudes for stars located within a radius 9$\arcmin$ 
from the respective cluster centres. We limited our sample to stars with proper motion errors $\le$ 0.1 
mas/yr, following the suggestion by \citet[][and reference therein]{piatti2021a}, who found 
that the larger  the individual proper motion errors of SMC stars, the larger the derived errors of the 
mean SMC
cluster proper motions. We minimized the presence of foreground stars by favoring distant stars 
(i.e., $|\varpi|$/$\sigma(\varpi)$ $<$ 3) and pruned the data with
\texttt{sepsi} $<$ 2 and \texttt{epsi} $<$ 1, which define a good balance between data quality and the 
number of retained objects for our sample \citep[see also][]{ripepietal2018}.
Once we applied the cleaning procedure to the {\it Gaia} EDR3 data sets, some star clusters
were left with few or no stars, because of not enough photometric depth. 

\section{Data availability}

Data used in this work are publicly available at https://datalab.noao.edu/smash/smash.php
site.

\section*{Acknowledgements}
I thank the referee for the thorough reading of the manuscript and
 suggestions to improve it. 

This research uses services or data provided by the Astro Data Lab at NSF's National Optical-Infrared Astronomy 
Research Laboratory. NSF's OIR Lab is operated by the Association of Universities for Research in Astronomy (AURA),
 Inc. under a cooperative agreement with the National Science Foundation.

This work has made use of data from the European Space Agency (ESA) mission
{\it Gaia} (\url{https://www.cosmos.esa.int/gaia}), processed by the {\it Gaia}
Data Processing and Analysis Consortium (DPAC,
\url{https://www.cosmos.esa.int/web/gaia/dpac/consortium}). Funding for the DPAC
has been provided by national institutions, in particular the institutions
participating in the {\it Gaia} Multilateral Agreement.



%\bibliographystyle{mnras}
%\bibliography{paper} % if your bibtex file is called paper.bib

%to be uncommented before sending to editor
%\input{paper.bbl}

%\begin{thebibliography}{99}
%\end{thebibliography}

%%%%%%%%%%%%%%%%%%%%%%%%%%%%%%%%%%%%%%%%%%%%%%%%%%
%%%%%%%%%%%%%%%% APPENDICES %%%%%%%%%%%%%%%%%%%%%

%\appendix


%%%%%%%%%%%%%%%%%%%%%%%%%%%%%%%%%%%%%%%%%%%%%%%%%%

% Don't change these lines
\bsp	% typesetting comment
\label{lastpage}
\end{document}